\documentclass{pasj01}
\usepackage{dcolumn}
\newcolumntype{d}[1]{D{.}{.}{#1}}
\draft 
\Received{$\langle$reception date$\rangle$}
\Accepted{$\langle$acception date$\rangle$}
\Published{$\langle$publication date$\rangle$}

\begin{document}

\title{An optical search for transients lasting a few seconds}

\author{Michael \textsc{W. Richmond}\altaffilmark{1}}
\author{Masaomi \textsc{Tanaka}\altaffilmark{2}}
\author{Tomoki \textsc{Morokuma}\altaffilmark{3}}
\author{Shigeyuki \textsc{Sako}\altaffilmark{3}}
\author{Ryou \textsc{Ohsawa}\altaffilmark{3}}
\author{Noriaki \textsc{Arima}\altaffilmark{3}}
\author{Nozomu \textsc{Tominaga}\altaffilmark{4,5}}
\author{Mamoru \textsc{Doi}\altaffilmark{1,6}}
\author{Tsutomu \textsc{Aoki}\altaffilmark{7}}
\author{Ko \textsc{Arimatsu}\altaffilmark{8}}
\author{Makoto \textsc{Ichiki}\altaffilmark{3}}
\author{Shiro \textsc{Ikeda}\altaffilmark{9}}
\author{Yoshifusa \textsc{Ita}\altaffilmark{2}}
\author{Toshihiro \textsc{Kasuga}\altaffilmark{10,11}}
\author{Koji S. \textsc{Kawabata}\altaffilmark{12}}
\author{Hideyo \textsc{Kawakita}\altaffilmark{11}}
\author{Naoto \textsc{Kobayashi}\altaffilmark{3,7}}
\author{Mitsuru \textsc{Kokubo}\altaffilmark{2}}
\author{Masahiro \textsc{Konishi}\altaffilmark{3}}
\author{Hiroyuki \textsc{Maehara}\altaffilmark{13}}
\author{Hiroyuki \textsc{Mito}\altaffilmark{7}}
\author{Takashi \textsc{Miyata}\altaffilmark{3}}
\author{Yuki \textsc{Mori}\altaffilmark{7}}
\author{Mikio \textsc{Morii}\altaffilmark{9}}
\author{Kentaro \textsc{Motohara}\altaffilmark{3}}
\author{Yoshikazu \textsc{Nakada}\altaffilmark{3}}
\author{Shin-Ichiro \textsc{Okumura}\altaffilmark{14}}
\author{Hiroki \textsc{Onozato}\altaffilmark{15}}
\author{Yuki \textsc{Sarugaku}\altaffilmark{11}}
\author{Mikiya \textsc{Sato}\altaffilmark{sato}}
\author{Toshikazu \textsc{Shigeyama}\altaffilmark{6}}
\author{Takao \textsc{Soyano}\altaffilmark{7}}
\author{Hidenori \textsc{Takahashi}\altaffilmark{7,3}}
\author{Ataru \textsc{Tanikawa}\altaffilmark{16}}
\author{Ken'ichi \textsc{Tarusawa}\altaffilmark{7}}
\author{Seitaro \textsc{Urakawa}\altaffilmark{14}}
\author{Fumihiko \textsc{Usui}\altaffilmark{18}}
\author{Junichi \textsc{Watanabe}\altaffilmark{10}}
\author{Takuya \textsc{Yamashita}\altaffilmark{10}}
\author{Makoto \textsc{Yoshikawa}\altaffilmark{19}}
\altaffiltext{1}{Physics Department, Rochester Institute of Technology, 84 Lomb Memorial Drive, Rochester, NY, 14623 USA}
\altaffiltext{2}{Tohoku University, 6-3 Aramaki, Aoba, Aoba-ku, Sendai, Miyagi 980-8578, Japan}
\altaffiltext{3}{Institute of Astronomy, Graduate School of Science, University of Tokyo, 2-21-1, Osawa, Mitaka, Tokyo 181-0015, Japan}
\altaffiltext{4}{Department of Physics, Faculty of Science and Engineering, Konan University, 8-9-1 Okamoto, Kobe, Hyogo 658-8501, Japan}
\altaffiltext{5}{Kavli Institute for the Physics and Mathematics of the Universe (WPI), The University of Tokyo, 5-1-5 Kashiwanoha, Kashiwa, Chiba 277-8583, Japan}
\altaffiltext{6}{Research Center for the Early Universe, Graduate School of Science, The University of Tokyo, 7-3-1 Hongo, Bunkyo-ku, Tokyo 113-0033, Japan}
\altaffiltext{7}{Kiso Observatory, Institute of Astronomy, School of Science, The University of Tokyo 10762-30, Mitake, Kiso-machi, Kiso-gun, Nagano 397-0101, Japan}
\altaffiltext{8}{Astronomical Observatory, Graduate School of Science,  Kyoto University, Kitashirakawa-oiwake-cho, Sakyo-ku, Kyoto 606-8502, Japan}
\altaffiltext{9}{The Institute of Statistical Mathematics, 10-3 Midori-cho, Tachikawa, Tokyo 190-8562, Japan}
\altaffiltext{10}{National Astronomical Observatory of Japan, 2-21-1 Osawa Mitaka Tokyo 181-8588 Japan}
\altaffiltext{11}{Department of Physics, Kyoto Sangyo University, Motoyama Kamigamo Kita-ku Kyoto 603-8555 Japan}
\altaffiltext{12}{Hiroshima Astrophysical Science Center, Hiroshima University, Higashi-Hiroshima, Hiroshima 739-8526, Japan}
\altaffiltext{13}{Okayama Branch Office, Subaru Telescope, National Astronomical Observatory of Japan, NINS, Kamogata, Asakuchi, Okayama, Japan}
\altaffiltext{14}{Japan Spaceguard Association, Bisei Spaceguard Center, 1716-3 Okura, Bisei, Ibara, Okayama 714-1411, Japan}
\altaffiltext{15}{Nishi-Harima Astronomical Observatory, Center for Astronomy, Institute of Natural and Environmental Sciences, University of Hyogo, 407-2, Nishigaichi, Sayo-cho, Sayo-gun, Hyogo 679-5313, Japan}
\altaffiltext{16}{Kawasaki Municipal Science Museum, 7-1-2 Masugata, Tama-ku, Kawasaki, Kanagawa, 214-0032, Japan}
\altaffiltext{17}{College of Arts and Sciences, the University of Tokyo, 3-8-1 Komaba, Meguro-ku, Tokyo 153- 8902, Japan}
\altaffiltext{18}{Center for Planetary Science, Graduate School of Science, Kobe University, 7-1-48 Minatojima- Minamimachi, Chuo-Ku, Kobe, Hyogo 650-0047, Japan}
\altaffiltext{19}{Japan Aerospace eXploration Agency, 3-1-1 Yoshinodai, Chuo-ku, Sagamihara, Kanagawa 252- 5210, Japan}


\maketitle

\begin{abstract}
Using a prototype of the Tomo-e Gozen wide-field CMOS mosaic camera,
we acquire wide-field optical images at a cadence of 2 Hz
and search them for transient sources of duration 1.5 to 11.5 seconds.
Over the course of eight nights, 
our survey encompasses the equivalent of roughly two days on
one square degree,
to a fluence equivalent to a limiting magnitude about $V = 15.6$
in a 1-second exposure.
After examining by eye the candidates identified by a software pipeline,
we find no sources which meet all our criteria.
We compute upper limits to the rate of optical transients
consistent with our survey,
and compare those to the rates expected and observed
for representative sources of ephemeral optical light.
\end{abstract}

\section{Introduction}


The sky is filled with objects which appear eternal
to the average human eye.  
True, the position and brightness of the planets 
change from week to week,
and that of the Moon from night to night,
but every generation of humans has seen
the same Moon.
To ancient skywatchers,
the fixed stars seemed even more constant.
On the rare occasions when a comet might appear
for a few months,
its ephemeral nature was held as a reason that
it must be an atmospheric phenomenon and not
a true member of the celestial sphere.

Over the past few centuries,
astronomers have recognized that some stars do vary
appreciably in brightness,
and all drift across the sky, albeit slowly.
Systematic surveys starting in the
nineteenth century found evidence for
stellar variation on timescales of years,
days, and even hours.
In recent decades, improving technology in
detectors, telescope control systems,
and data analysis has led to the creation of larger
and larger catalogs of variable stars.
Astronomers have also 
detected many
transient objects:
novae, supernovae, gamma-ray bursts (GRBs),
fast radio bursts,
and tidal disruption events,
to name just a few.

Despite these efforts,
some regions of parameter space remain largely unexplored.
In particular, objects which appear only 
once for durations of just a few seconds --
in a word, ``flashes'' --
have escaped serious study.
The compilation of studies presented by
\cite{2013ApJ...779...18B},
for example, 
includes no projects sensitive to flashes
shorter than 20 minutes.
There are at least three factors which make it difficult
to look for sources which emit for only a few minutes,
let alone a few seconds.
First, some detectors are not designed to operate at high 
cadence; large format CCDs require long readout times to 
produce images with low noise.
Second, in order to eliminate false positives, it is 
beneficial to acquire at least two, if not three,
measurements of a single event before it disappears;
yet it is difficult to accumulate a reasonable signal-to-noise
ratio in such short exposures.
Finally, long stretches of high-cadence measurements will
produce a very large dataset,
increasing the time and effort required to identify 
and confirm real transients.

Our effort was inspired by the development of 
large CMOS detectors which yield low noise
even when read out several times per second.
Placing a mosaic of such devices onto the focal
plane of a Schmidt telescope creates an instrument
which can sample large areas of the sky with
high cadence,
and designing a pipeline to analyze the
large volume of data carefully 
reduces the amount of human oversight 
required to detect transient sources reliably.
In section
{\ref{sec:obs}},
we describe our equipment and the methods
used to acquire images.
The processing of those images into calibrated
lists of stars is explained in section
{\ref{sec:proc}}.
We discuss the criteria
by which transient objects are 
chosen from the star lists in 
Section
{\ref{sec:search}},
and conclude that no objects satisfy all our conditions.
In section
{\ref{sec:rate}},
we use our null result to compute 
upper limits on the rates of transient sources
per square degree.
We explain how to convert the rates into volumetric units in section
{\ref{sec:volume}},
and briefly compare them to observed and expected rates
from the literature
for two types of source.
We compare our survey to three others which
sample timescales of a few seconds
in section
{\ref{sec:comparison}}.

We assume throughout this work a cosmology with
$H_0 = 69.6$ km/s/Mpc, $\Omega_M = 0.286$ and $\Omega_{\Lambda} = 0.714$.

\section{Observations}
\label{sec:obs}

All images were acquired with the 105-cm Schmidt Telescope
at Kiso Observatory,
at 
$35^\circ 47{'} 50{\rlap.}{''}0$ North,
$137^\circ 37' 31{\rlap.}{''}5$ East in the WGS84 system, 
and an altitude of 1132 m.
Light was focused on an early version of the Tomo-e Gozen camera
\citep{2016SPIE.9908E..3PS, 2018SPIE10702E..0JS},
referred to hereafter as Tomo-e PM (for ``Prototype Model'').
This prototype had 8 CMOS sensors installed on the focal plane,
each with $2000 \times 1128$ pixels of size $19\ \mu$m on a side.
The plate scale was approximately $1{\rlap.}{''}19$ arcseconds
per pixel.
Each chip subtended an area of approximately
$39.7 \times 22.4$ arcminutes,
and the set of eight covered a total of about 1.9 square degrees,
aligned in a row running east to west.
Image frames were obtained by a rolling read operation 
at two frames per second;
since the pixel reset time is only 0.1 milliseconds, 
the exposure time of each frame was approximately 0.5 seconds.
No filters were placed between the sensors and the sky,
so the detectors measured light over the range
$370$ to $730$ nm, with a peak quantum efficiency
of $0.68$ at $\lambda = 500$ nm.

\begin{figure}
   \begin{center}
      \includegraphics[width=80mm]{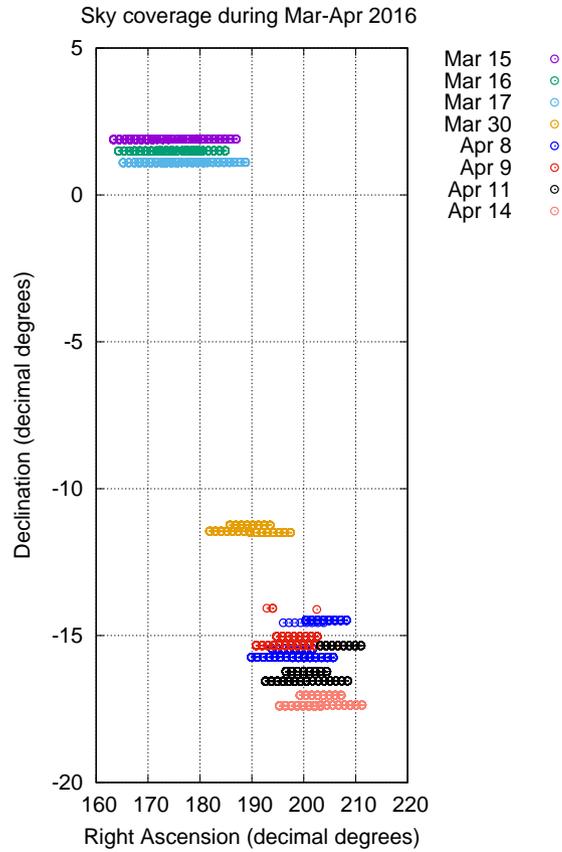}
   \end{center}
    \caption{Sky coverage during eight nights in March-April, 2016.
              A different color is used for each night.
              Each symbol shows the center of a field, but the size
              of the symbols is not equal to the area of each field.}
    \label{fig:radec}
\end{figure}

The data analyzed herein were gathered during eight nights
in March and April of 2016;
see Table 
\ref{tab:nights}
for a list of the nights.
In order to minimize the number of artificial satellites 
and space debris in our images,
we aimed the telescope in the direction
of the the Earth's shadow.
During each night, the pointing of the telescope gradually moved 
west to follow the shadow;
figure 
\ref{fig:radec}
shows the area covered during each night.
The center of this region is approximately at
${\rm RA} = 190^{\circ}$, ${\rm Dec} = -8^{\circ}$,
which corresponds to galactic coordinates
${\rm \ell} = 298^{\circ}$, ${b} = +55^{\circ}$
and ecliptic 
${\rm long} = 192^{\circ}$, ${\rm lat} = -3^{\circ}$.
It is far from the plane of the Milky Way,
but close to the plane of the ecliptic.

\begin{table}
  \caption{Nights of observation}
  \label{tab:nights}
  \begin{tabular}{llll}
    \hfil Date \hfil & hours & FWHM (arcsec) & weather \\
    \hline
    2016 March 15  &  8.4  &  $4 - 5$  &  good \\
    2016 March 16  &  6.7  &  $3 - 5$  & brief clouds \\
    2016 March 17  &  7.9  &  $3 - 4$  & mediocre \\
    2016 March 30  &  5.0  &  $4 - 7$  & mediocre \\
    2016 April 08  &  6.5  &  $4 - 6$  & good \\
    2016 April 09  &  5.3  &  $3 - 4$  & light clouds \\
    2016 April 11  &  6.5  &  $5 - 8$  & good \\
    2016 April 14  &  6.2  &  $5 - 8$  & light clouds \\
  \end{tabular}
\end{table}

\section{Processing the images}
\label{sec:proc}

The real-time electronics collect the information from 
each sensor and package it into units we call ``chunks,''
which consist of 360 individual images
(180 seconds of data at 2 Hz).
As described below, we sometimes treat the
information from a single chunk as a unit for 
calibration.

We list below the steps involved in converting each
raw image from a single sensor into a list of 
astrometrically and photometrically calibrated objects.

\begin{enumerate}
  \item Following the method described in
              \cite{2016SPIE.9913E..39O},
              we create a simple model of the bias
              for each image; it has a structure which repeats
              every four rows.
              We subtract this bias from the raw image pixels.

   \item We do {\it not} perform any flatfielding.

   \item Using the {\tt stars} and {\tt phot} routines of the
              XVista package
              \citep{1989PASP..101..725T, RichmondXVista},
              we identify objects with peaks significantly above 
              the sky level and perform aperture photometry.
              Objects with sizes much smaller or larger than 
              the typical stellar FWHM are excluded.
              We measure the flux of each object within circular
              apertures of two sizes: one fixed at 10 pixels 
              ($11{\rlap.}{''}8$) 
              and one which varied from image to image,
              equal to $1.5$ times the FWHM of the image.
              The result of this stage is one list of stars
              per image, with (row, col) positions and 
              instrumental magnitudes on an arbitrary scale.

    \item Objects detected at least twice in the images of a single chunk are
              combined into an ensemble.  These objects are 
              subjected to inhomogeneous ensemble photometry
              \citep{1992PASP..104..435H},
              bringing them to a single photometric zeropoint.
              This procedure creates a set of robust statistics,
              such as the photometric uncertainty as a function
              of instrumental magnitude.

    \item We calibrate the ensemble averages of each object 
              astrometrically against the UCAC4
              \citep{2013AJ....145...44Z},
              and photometrically against the UCAC4's $V-$band
              magnitudes;
              recall that our measurements are unfiltered.
              This provides accurate measurements for stars 
              which appear in most of the images of the chunk;
              in other words, for ordinary stars which do not vary
              on short timescales.

    \item The objects detected in each individual image are
              also calibrated against the UCAC4, astrometrically
              and photometrically, in a similar manner.
              These results will be less accurate, but provide
              measurements for transient objects which might not
              be included in the ensembles.
             
    \item We create a set of diagnostic graphs and tables, summarizing the 
              properties of objects in each chunk.
            
\end{enumerate}

We employed a simple first-order model to match the (x, y) positions
of stars on each sensor to the (RA, Dec) positions from the 
UCAC4 catalog.  The typical solution involved 30 to 50 stars
and yielded residuals of size 
$\sim 0{\rlap.}{''}5$, 
adequate for our purposes.
See Figure
{\ref{fig:astplot} for details.

\begin{figure}
   \begin{center}
      \includegraphics[width=80mm]{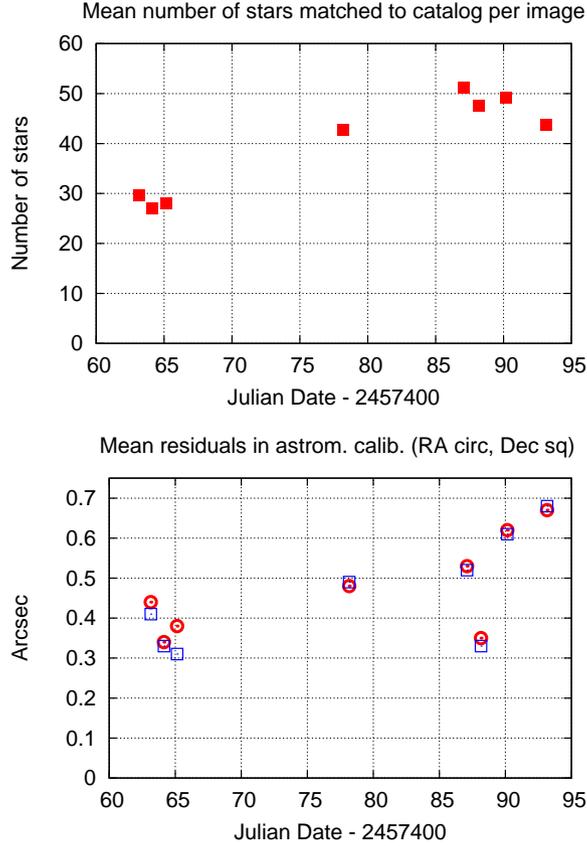}
   \end{center}
    \caption{Properties of astrometric solutions in March and April, 2016.
             In the lower panel, mean residuals in RA are shown as red circles,
             in Dec by blue squares.}
    \label{fig:astplot}
\end{figure}

We found that using photometric apertures which varied with the seeing
produced better results than using fixed apertures,
not surprising given the considerable variations in FWHM
over the course of some of our nights.
Depending on the conditions,
bright stars ($V \sim 11$) showed scatter of 0.02 to 0.03 mag,
while measurements of faint stars ($V \sim 16$) 
varied by 0.08 to 0.12 mag from 
image to image.
Figure
{\ref{fig:sigmaplot} 
shows examples under good and adverse conditions.
Note that our very short exposures mean that scintillation noise
can contribute significantly to the noise floor
at moderate airmass.

\begin{figure}
   \begin{center}
      \includegraphics[width=80mm]{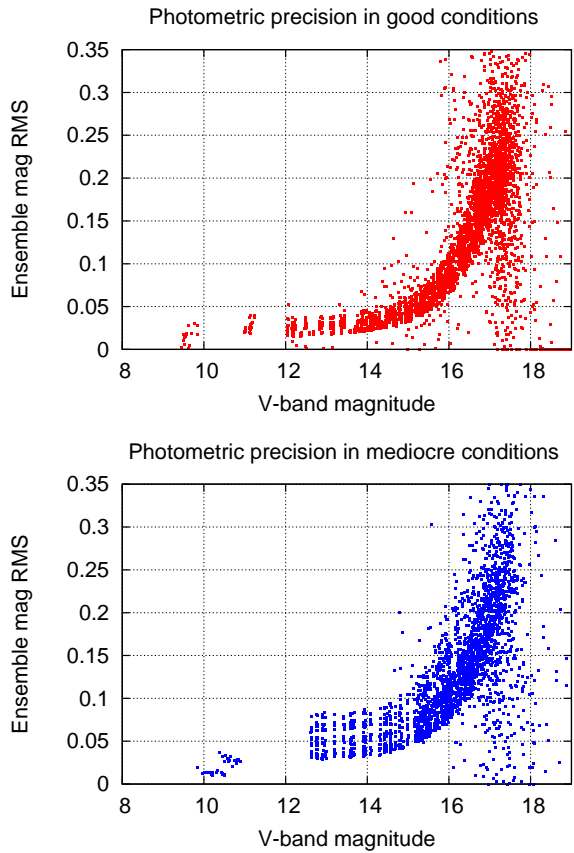}
   \end{center}
    \caption{Scatter in repeated measurements as a function of magnitude.
             Top panel: good conditions, 2016 Mar 15 around JD 2457463.06;
             bottom panel: poor conditions, 2016 Mar 17 around JD 2457465.06}
    \label{fig:sigmaplot}
\end{figure}

The pipeline produces two lists of stars
as its main output.
One is based on the ensemble of stars within each chunk
of images;
the other is a list of every object detected in each individual
image.
Each list is calibrated in position and magnitude
against the UCAC4 catalog.
The pipeline also produces a set of diagnostic graphs
to help us determine the quality of each night.

%
%

\section{Search for transients}
\label{sec:search}

Running the pipeline takes a significant amount of computing
power and time.
We designed our algorithm which searches for transient objects
to operate on the star lists created by the pipeline.
Since the data volume is much smaller (by factors of $\sim 100$),
the time to search, check, and if necessary re-search
is much smaller.

Our method is based on measurements made within a single
chunk of data: 360 consecutive images spanning 180 seconds.
As described below, this can cause the method 
to miss a very small fraction of very brief transients;
however, we accept this small drawback in return for a significant
simplication in the bookkeeping.

We scan through all objects detected within the chunk
and subject them to a series of tests.
An object which fails any test is discarded.
In order to pass these tests, 

\begin{enumerate}
  \item an object must be detected between 3 and 20 times within 
              a single chunk of 360 images.

  \item all detections must appear within a window of 
              $(N + 3)$ consecutive images, where
              $N$ is the number of detections.
              In other words, we look only for objects which
              brighten once, then fade away;
              we discard objects which appear repeatedly at regular
              or irregular intervals.  
              We add $3$ extra images to our window of acceptance
              to accomodate objects close to the limiting magnitude;
              small random variations in transparency or seeing
              may cause them to fall just below the detection 
              threshold in, say, one of six images.
              Figure {\ref{fig:criteria}} shows example
              sequences of detections and non-detections,
              and our decision in each case.
              In the terms used by 
              \cite{2019arXiv190311083A},
              our criteria are {\tt min=3, max=20, holes=3, empty=0}.

\begin{figure}
   \begin{center}
      \includegraphics[width=80mm]{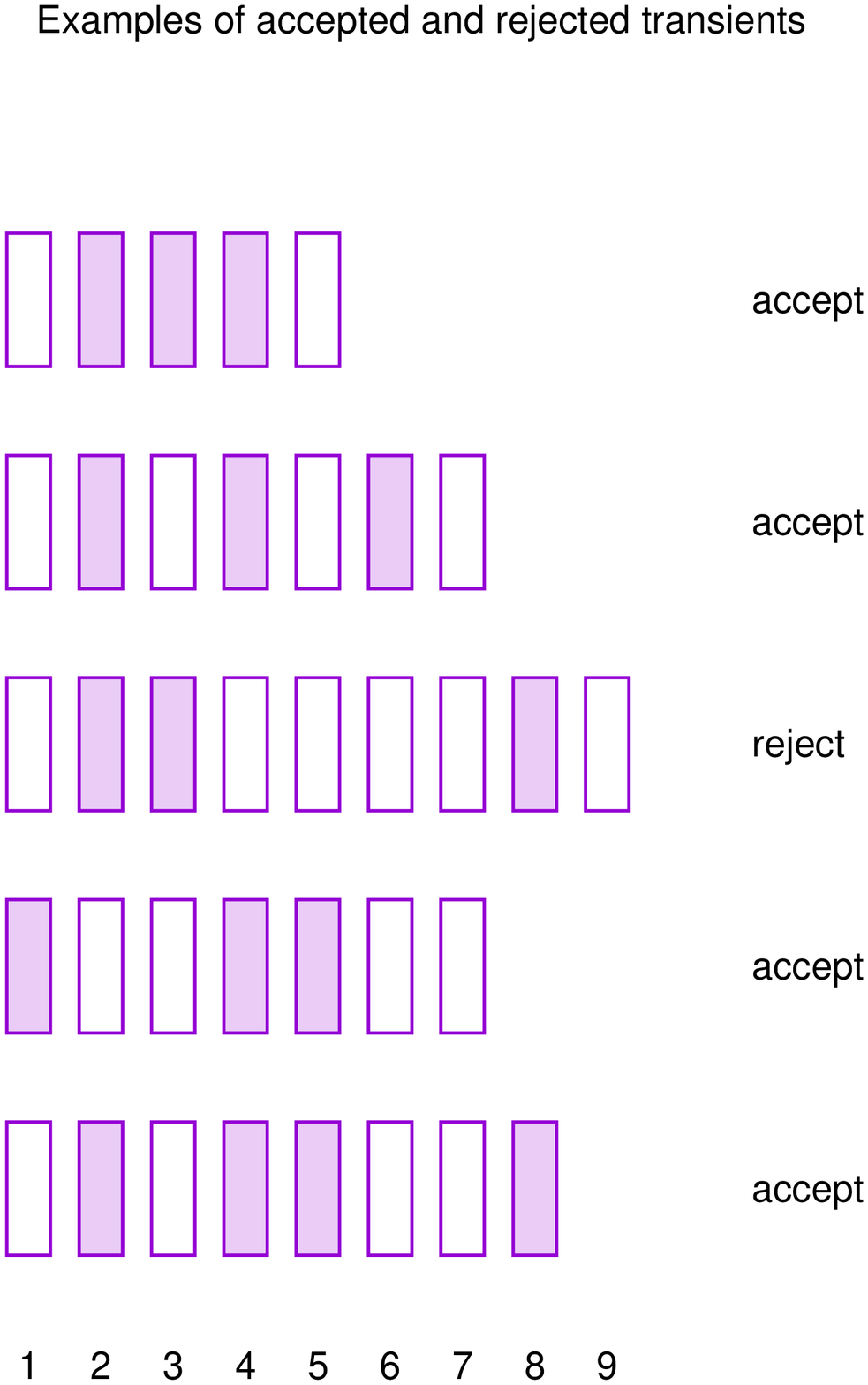}
   \end{center}
    \caption{Examples of sequences of images with detections
             (filled rectangles) and non-detections (empty rectangles),
             and our algorithm's decision at right.}
    \label{fig:criteria}
\end{figure}

   \item the magnitude of a candidate must be less 
              than $(m + 1)$, where $m$ is the limiting magnitude
              of the chunk (which we will discuss below).
              In other words, the object may be slightly fainter,
              on average, than the limiting magnitude.
              
   \item the object must be at least 7 pixels ($8\rlap{.}{''}3$)
              from its nearest neighbor in the ensemble.
              This rule rejects many bogus candidates in the
              outer regions of bright stars.

   \item the object must be at least 10 pixels ($11\rlap{.}{''}8$)
              from the edges of the image.
              Our routines for measuring accurately the properties
              of an object set this limit.

\end{enumerate}
    
Conditions 1 and 2 can cause a real, very brief transient to 
be missed.
If an object appears for exactly 3 images (1.5 seconds),
and one of those images happens to be either the first or last
frame in a chunk,
or if it appears for exactly 4 images (2 seconds),
and two of those happen to be the first two or last two 
of a chunk,
then it will not pass the tests.
Since this applies only to a very small fraction
of all possible transients (about $0.6\%$ of those lasting
with exactly 3 or 4 detections),
we accept the losses.

Condition 5 causes us to ignore a thin strip around the
edges of each image,
amounting to $2.7\%$ of the area of the survey.
We will account for this small inefficiency in our calculations
of the control time and transient rates 
later in this paper.

Describing the ``limiting magnitude'' for an instrument
can be approached in several ways.
For example, 
under optimal conditions, a star of magnitude
$V = 18.5$ produces a signal-to-noise ratio of 5 
in 1-second exposure with Tomo-e
\citep{2018SPIE10702E..0JS};
so one might say that the 0.5-second exposures 
we studied ought to have a limiting magnitude 
of $V = 17.7$
However, since our search was often conducted in less than
optimal conditions, we take a more conservative approach
in this paper.
In order to calculate the limiting magnitude for a chunk,
we begin by counting the number $N$ of ``good'' images with
in the chunk.  
This is normally close to 360,
but may be smaller if some frames are rendered unusable
by clouds, bad seeing, passing airplanes, or other problems.
We then compare the number of images in which 
each star is detected to $N$.
Bright stars will appear in nearly all of them,
but very faint stars may rise above the noise threshold 
only in a few images.
We define the ``limiting magnitude'' of a chunk
to be the magnitude at which the average number
of detections falls to $N/2$.
%
%
We record this limiting magnitude for each chunk,
and use it not only to test possible transients,
but also to compute control times 
(see section 
{\ref{sec:rate}}).

After running all the star lists through these tests,
we found a total of 57 candidate transient objects.
We examined each candidate in detail, scrutinizing the
images before, during, and after its detections.
We discovered that most of these candidates
were simply ordinary stars, somewhat fainter than the
limiting magnitude of our observations.
Due to fortuitous moments of good seeing, they would 
sometimes pop up above the detection limit for a
few consecutive frames before disappearing below the sky
noise again.
Although our software failed to notice them
most of the time,
a human eye could often detect them, once their positions were known.
We examined 2MASS and Digitized Sky Survey images at the positions
of all candidates to look for faint stars,
and rejected all candidates with counterparts in these surveys.
One example of this sort of object is shown
in Figure
{\ref{fig:faintstar}}.

\begin{figure}
   \begin{center}
      \includegraphics[width=160mm]{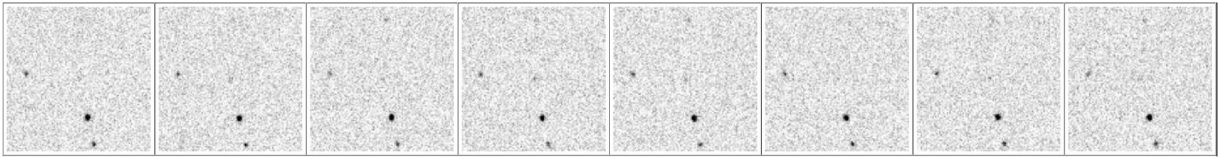}
   \end{center}
    \caption{Cutouts centered on a candidate of March 30, 2016.
                The object was detected in the second, fourth, 
                and seventh images in this sequence,
                but is present in all of them.
                Each cutout is 130 arcseconds on a side, 
                with North up, East left.
                Time runs from left to right.}
    \label{fig:faintstar}
\end{figure}

A few of the candidates were bright stars, far above the
detection limit,
which appeared offset from their usual positions
due to particularly bad seeing.  
Short-term atmospheric fluctuations can lead to 
unusually distorted point spread functions when
one integrates for very short durations.
An example of this type of candidate is shown
in Figure
{\ref{fig:seeing}}.

\begin{figure}
   \begin{center}
      \includegraphics[width=160mm]{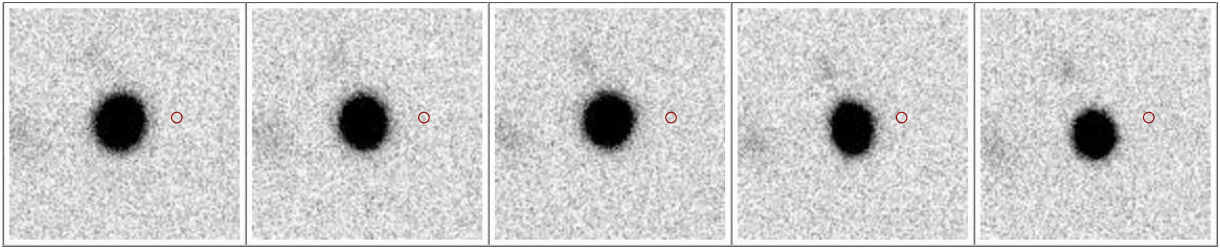}
   \end{center}
    \caption{Cutouts centered on a candidate of Apr 11, 2016,
              during conditions of poor seeing.
                The software determined a position for this object
                in the central 3 images which was sufficiently offset
                from the usual position that it appeared as a new object.
                The red circle indicates the size of matching radius.
                Each cutout is 130 arcseconds on a side, 
                with North up, East left.
                Time runs from left to right.}
    \label{fig:seeing}
\end{figure}

The most promising candidates appeared on the night of 
2016 March 17,
near RA $= 169.24$ and Dec $= 0.97$.
Two of the seven candidates for this night
turned out to be the same object,
which was moving at a rate of about 
$5^{''}$ per second to the East.
It was below the threshold of detection
for most of the observations, 
but rose above the threshold for a few seconds
and was detected twice,
separated by a few seconds.
Figure
{\ref{fig:candidate}}
shows subframes of consecutive images,
centered on the candidate.
The top row shows the first detection,
and the lower row the second detection.

\begin{figure}
   \begin{center}
      \includegraphics[width=160mm]{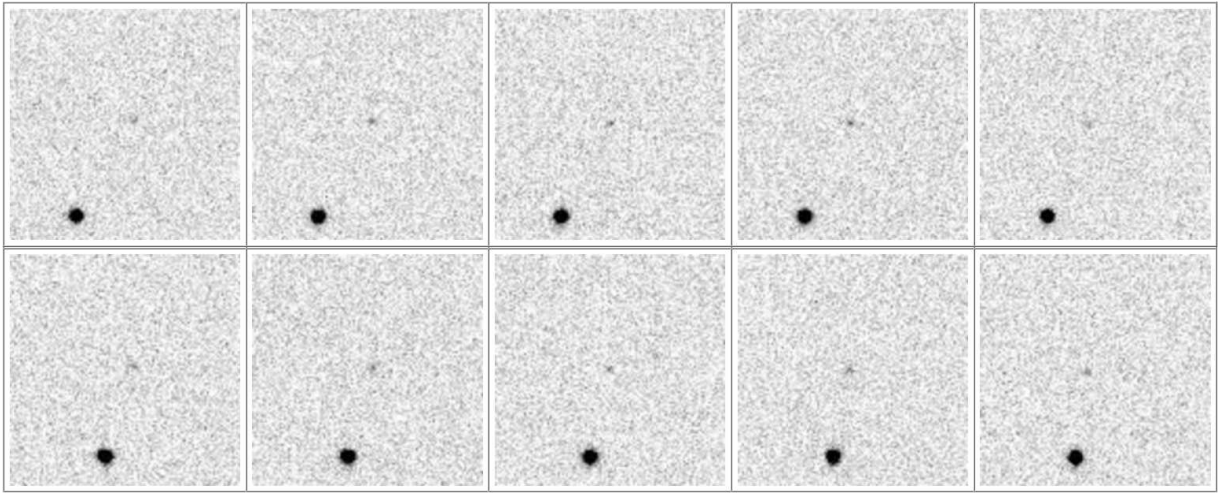}
   \end{center}
    \caption{Cutouts centered on the moving candidate from March 17, 2016.
                Each cutout is 130 arcseconds on a side, 
                with North up, East left.
                Time runs from left to right in each row.}
    \label{fig:candidate}
\end{figure}

Using tools provided by the Minor Planet Center,
we found no known asteroids at this position.
The motion is consistent with that of an artificial
satellite in high-Earth orbit.
When we examined the intervening images, 
we were able to detect it in most of them --
but at a significance level below that required
to trigger our star-finding routines.

In summary,
we found zero stationary objects which appeared, and then disappeared,
over a span of time between 1.5 and 11.5 seconds.
However, the fact that our procedure did uncover
a very faint, moving satellite,
at two locations along its orbit,
gives us some confidence that 
it would have flagged a real transient source
as well.

What can we learn from this null result?

\section{The rate of transient events}
\label{sec:rate}

Following methods used by supernova researchers
starting with
\cite{1938ApJ....88..529Z},
we introduce the concept of {\it control time}:
the time during which a transient event would have been
detected by our system, 
should it have occurred.
In the case of supernovae, which remain bright for
many weeks,
one must take into account the possibility
of an explosion happening some time before the start
of an observation.
Since we are interested in objects which fade only
a few seconds after their rise, however,
an interval much shorter than the duration of our measurements
each night,
our control time is simply the sum of our
exposure times
(after removing periods ruined by clouds, bad seeing,
and other problems).

In the decades following
\cite{1938ApJ....88..529Z},
survey images typically covered the entire
extent of some particular galaxy
(and of many galaxies, in fact).
Any supernova in that galaxy would appear in each
survey image,
and so the control time for that galaxy 
would be simply a function of the duration of the
survey and the intervals between images.
In most modern searches, including ours,
which are not confined to some known and finite source,
the control time must include a factor
in addition to time: the area over which new objects could be
detected.
Clearly, for example,
an eight-hour sequence of images covering ten square degrees
provides more information than an eight-hour sequence of 
images reaching the same depth, but covering only one square degree.
This example makes clear that another variable 
we must consider is the
depth of the survey:
an eight-hour sequence of images reaching $V=15$ provides
more information than a similar sequence of images
which reaches only $V=10$.

In order to characterize our survey for transients,
therefore, 
we must describe 

\begin{itemize}

  \item{the area of the sky covered; we will use square degrees}

  \item{the time span of the observations; we choose seconds}

  \item{the depth of the survey; see discussion below}

\end{itemize}

In most circumstances,
astronomers would describe the depth of a survey
by its limiting magnitude, or limiting flux;
these are appropriate choices to describe sources 
which are constant in brightness.
Since we are searching for objects which might change in
brightness considerably from one image to the next,
however,
or even during the exposure of a single image,
it makes more sense to consider the concept of {\it fluence}.
Fluence is the integral of flux over time;
it might be the number of photons collected during
one exposure, for example, or summed over several exposures.

Since our search criteria require that an object be detected
in a series of (nearly) consecutive images,
we can compute a rough lower limit on the fluence of an
object that our algorithm would find:

$$
{\rm Fluence\ limit\ F\ } = 10^{-0.4 m} \thinspace f t N
$$

\noindent 
where
$m$ is the ordinary limiting magnitude of a single image,
$f$ is the flux density corresponding to an object of magnitude zero,
$t = 0.5$ seconds is the exposure time of one of our images,
and 
$N = 3$ is the minimum number of consecutive images
in which an object must appear.
We adopt a value of 
$f = 3.16 \times 10^{-6} \thinspace {\rm erg/cm^{2} s}$,
corresponding to the $V$-band zeropoint.

As described in the previous section,
our pipeline computes a limiting magnitude $m$ for each
chunk of images.
We apply that value to all the images in a chunk
in order to compute control times.
Thus, we can not account properly for 
any changes in sky conditions on timescales
shorter than the duration of a chunk (180 seconds).

Even though fluence is a better choice when 
describing the behavior of very brief transients,
most of the literature on surveys for supernova,
gamma-ray bursts, and other variable objects
characterizes their measurements in terms
of a limiting magnitude.
In order to compare our results more easily with
others, 
we will convert our results into those terms.
Since we are looking for objects with lifetimes
of just a few seconds,
it is convenient to describe our results
in terms of
the fluence over a period of one second,
or, equivalently,
{\it the limiting magnitude in a 1-second exposure}.
Since the conditions changed significantly within
some nights, as well as from night to night, 
and since the properties of the sources which might 
appear in our data have very broad ranges,
we have chosen to compute control times for a range
of observed fluences, 
rather than a single value.
We compute control times for four choices 
of this parameter,
corresponding to limiting magnitudes of
$V = 12.6, \ 13.6, \ 14.6, {\rm \ and\ } 15.6$, in a one-second exposure.`
The results are presented
in Table
{\ref{tab:controltimes}.
The small value in the final entry for 2016 Apr 14
is due to a combination of bad seeing and clouds.

\begin{table}
  \caption{Control time (sq.deg.*sec) for several thresholds}
  \label{tab:controltimes}
  \begin{tabular}{lllll}
    \hfil Date \hfil & $V=12.6$ & $V=13.6$ & $V=14.6$ & $V=15.6$ \\
    \hline
    2016 March 15  &  49,357  &  49,357  &  49,357  &  43,685 \\

    2016 March 16  &  40,520  &  40,520  &  38,834  &  33,083  \\

    2016 March 17  &  43,984  &  43,741  &  41,077  &  32,671 \\

    2016 March 30  &  30,133  & 29,292   &  25,108  &  19,253 \\

    2016 Apr 08    &  33,068  &  31,919  &  28,311  &  23,028 \\

    2016 Apr 09    &  30,015  &  29,298  &  25,848  &  10,807 \\

    2016 Apr 11    &  36,771  &  34,252  &  28,888  &  14,843 \\

    2016 Apr 14    &  28,216  &  25,715  &  20,015  &     128 \\
    \hline

     total         &  292,069  & 284,100 &  257,441 &  177,502 \\

  \end{tabular}
\end{table}

Our goal is to use this information to determine
the rate at which transients appear,
in units of objects per square degree per second.
Since we detected no events, however,
we cannot perform this calculation in the simple way.
Instead, we will use our measurements to place an
upper limit on the rate at which transients
appear.
We will begin by making two assumptions:

\begin{enumerate}

  \item transient events are independent of each other

  \item the rate of events, $\lambda$, is constant during the course
             of our survey 

\end{enumerate}

\noindent
With these assumptions,
we can model the occurence of events as a Poisson distribution,
so that the probability $p$ of $N$ events during
some period $t$ is

$$
p = {{ (\lambda t)^N \thinspace e^{- \lambda t}} \over {N!}}
$$

We observed $N = 0$ events.
In order to place limits on the rate $\lambda$,
we must choose some small probability $p_0$ 
that this lack of detections is due purely to chance.
We make the arbitrary choice of $p_0 = 0.05$,
and then reject any rate $\lambda$ 
which would yield a probability $p < p_0$
of observing zero events.
This technique yields upper limits on the rate
of transient events meeting our criteria,
which we quote
in units of events per square degree per day.
These values are listed in Table
{\ref{tab:rates}}
for the four choices of limiting fluence,
and shown in
Figure
{\ref{fig:arealrates}}.

\begin{table}
  \caption{Upper limit (95\% confidence level) on rate (events per sq.deg. per day) for several thresholds}
  \label{tab:rates}
  \begin{tabular}{lllll}
    \hfil Fluence\footnotemark[$*$] \hfil & $V=12.6$ & $V=13.6$ & $V=14.6$ & $V=15.6$ \\
    \hline
     Rate         & $0.89$ & $0.91$ &  $1.00$ &  $1.46$ \\

  \end{tabular}
  \begin{tabnote}
     \footnotemark[$*$] Equivalent limiting mag in 1-sec exposure.
  \end{tabnote}
\end{table}

\begin{figure}
   \begin{center}
      \includegraphics[width=80mm]{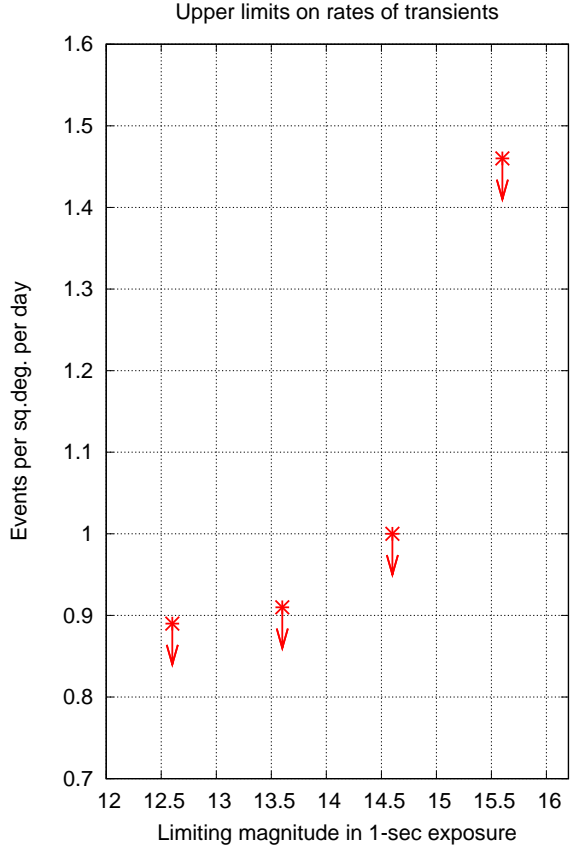}
   \end{center}
    \caption{Upper limits to the areal rates of optical flashes}
    \label{fig:arealrates}
\end{figure}

\section{The volumetric rate of specific sources}
\label{sec:volume}

If we know the properties of some specific type
of brief-lived source,
we can convert our limits from area-based
(per square degree per second)
to volume-based (per cubic pc per second).
Let us assume that an explosion of some type
produces a total energy $E$ 
in optical light 
over some short duration $\Delta t$ 
which lies within the range of our transient search.
Assuming that the brightness of the source
is roughly constant over this period,
its flux $f$ will be

$$
f = {{E}\over{4 \pi d^2} \thinspace \Delta t}
$$

For any one of our chosen flux limits 
$f_{\rm lim}$
in Table
{\ref{tab:rates}},
we can compute the maximum distance
out to which such a source would be detected 
by our survey:

$$
d_{\rm max} = \sqrt{  {{E}\over{4 \pi f_{\rm lim} \thinspace \Delta t}}  }
$$

To illustrate the method,
we choose two representive sources:
a ``nearby'' type,
the impact of a small body onto a white dwarf,
and a ``distant'' variety,
the prompt optical flash of a gamma-ray burst.
Our goal in the following discussion is
to show the steps one can follow to compute volumetric
rates from our survey's data.

\subsection{Asteroids and comets falling onto white dwarfs}

The collision of small bodies onto the surface of a 
white dwarf is discussed in
\cite{2015arXiv150107837D},
and one sees evidence for it in the abundance of some
heavy elements in the spectra of certain white
dwarfs.
The energy produced in the collision is released
over a period which increases with the mass of the impactor.
For objects of mass
$m \sim 10^{14}$ g,
\cite{2015arXiv150107837D}
estimate that the optical flash might last for 
$\Delta t \sim 1$ second,
and contain roughly
$E \sim 10^{31}$ ergs.
Using these values,
we compute distances out to which such collisions
would be detected,
and list them in Table
{\ref{tab:dist}}.

\begin{table}
  \caption{Detection limit $d_{\rm max}$ for $10^{31}$ erg/s comet-WD collision}
  \label{tab:dist}
  \begin{tabular}{lllll}
    \hfil source \hfil & $V=12.6$ & $V=13.6$ & $V=14.6$ & $V=15.6$ \\
    \hline
     comet-WD        &  54 pc &   86 pc  &  136 pc  & 215 pc \\
  \end{tabular}
\end{table}

Once we know the distance $d_{\rm max}$ 
out to which a source could be detected,
we can compute the effective volume
for that type of source covered by 
of our survey.
Let us perform one such calculation here as an example.
For a fluence limit of $V = 13.6$ mag in a 1-second exposure,
we could detect a comet-WD collision
out to a distance of $d_{\rm max} = 86$ pc.
Since our rates listed in 
Table
{\ref{tab:rates}
have units of ``per square degree,''
the effective volume of our search will be that of a 
narrow spherical sector with radius $d_{\rm max}$.
The ratio of the volume of this sector to the volume
of a full sphere will be the ratio of the area of
one square degree to the area of a full sphere;
thus,

$$
{\rm effective\ volume\ } V_{\rm eff} \quad = \quad 
         {{4 \pi}\over{3}} (d_{\rm max})^3
     \thinspace \left( {{1 {\rm \ sq.\ deg.}}\over{4 \pi \ {\rm ster}}}  \right)
  \quad = \quad 
     2.42 \times 10^{-5} \thinspace {{4 \pi}\over{3}} (d_{\rm max})^3
$$

In our example, since $d_{\rm max} = 86$ pc,
the effective volume is $V_{\rm eff} = 64.6$ cubic pc.
Multiplying this volume by the duration portion of the
control time will yield a ``control volume:''
in this case, 64.6 (cubic pc $\times$ day).
In a similar way,
we can convert our upper limit on the rate
of transients from areal to volumetric 
by dividing the rate in Table
{\ref{tab:rates}}
by this effective volume.
For our example, 

$$
{\rm volumetric\ rate\ } r_{\rm vol} \quad \leq  \quad
    {  {0.91 \thinspace {\rm events/sq.deg. \cdot day}}\over
         {64.6 \thinspace {\rm pc}^{-3}/{\rm sq.deg.}} }
   \quad = \quad 
         1.4 \times 10^{-2} \thinspace {\rm events}/{\rm pc^{-3} \cdot day}
$$

\cite{2015arXiv150107837D}
estimate that the number of comet-WD collisions
in a typical galaxy is $8 \times 10^{6}$ per year.
If we model the disk of the Milky Way as a thin
disk of radius $10$ kpc and height $1$ kpc,
its volume is $3 \times 10^{11} \ {\rm pc^{3}}$;
our upper limit to the rate (for a fluence equivalent
to $V = 15.6$ in a 1-second exposure)
would imply that no more than
$\sim 1.6 \times 10^{11}$ events should occur
within the entire galaxy each year.
While this is certainly consistent with the expections of 
\cite{2015arXiv150107837D},
it is far from being a strict constraint.

\subsection{Prompt optical flashes of gamma-ray bursts}

In many cases, optical emission has been detected 
hours to days after a GRB, as part of the physical
process called the afterglow.
In a very few cases, 
optical telescopes have seen flashes at the same time
as the initial burst of gamma rays.
We will 
take as models 
three events
in which the optical flash was particular bright:
GRBs 080319B 
\citep{2008Natur.455..183R, 2008AIPC.1059...59C, 2009ApJ...691..495W},
130427A
\citep{2013Natur.500..547T, 2014Sci...343...38V},
and 
160625B
\citep{2016GCN.19615....1B, 2017Natur.547..425T, 2017ASPC..510..309K}.
All three are classified as ``long'' GRBs, 
as instruments detected gamma rays from them for durations
longer than two seconds
(see 
\cite{2014ARA&A..52...43B}
and
\cite{2015AdAst2015E..22P}
for reviews of GRB properties).
In each of these cases,
the GRB seems to have been caused by the collapse
of a massive star.

The calculations of $d_{\rm max}$
and volumetric rate for these very luminous and distant
events are more complicated than those for ``nearby''
objects for two reasons.
First, the relationship between distance and apparent brightness
is not the simple Euclidean $1/r^2$, due to 
the curvature of space-time and relativistic effects.
Second, when determining the observed properties of objects at
redshifts $z > 0.5$,
one must account for the difference between the spectral
region in which light was emitted in the rest frame,
and the spectral region in which the light is detected
by the observer;
in other words, one must know the shape of the source's spectral
energy distribution.

Based on the optical and gamma-ray observations
of the 3 model GRBs in our comparison set,
we have constructed a ``toy model'' of the prompt
emission from such an event:
a simple power-law spectrum with a cutoff at
$h \nu_0 = 2000$ keV.

$$
{\rm flux\ density\ } f_\nu = \left\{
\begin{array}{ll}
   A \thinspace {\left( {{\nu}\over{\nu_0}} \right)}^{-0.5} & \nu \leq \nu_0 \\
   0 & \nu > \nu_0 \\
\end{array} 
\right.
$$

The amplitude $A$ is proportional to the overall 
luminosity of prompt emission.
We choose to describe it by computing an
``overall gamma-ray peak luminosity,'' 
$L_\gamma$,
for the flash, 
which we calculate using rest-frame emission
in the range 
$10 - 2000$ keV.
For a wide range of such luminosities,
we then determine the redshifts at which
our model's observed optical brightness
would fall below our four limiting magnitudes;
the results are shown in Figure
{\ref{fig:limitz}}.
Note that the intergalactic medium is likely to 
be ionized at 
$z \gtrsim 5.5$
\citep{2004AJ....127.2598S},
and hence opaque to the (at that time)
ultraviolet photons flying toward us.
However, it is possible that some windows of 
low opacity might allow the detection of
a small fraction of sources beyond this redshift.

\begin{figure}
   \begin{center}
      \includegraphics[width=80mm]{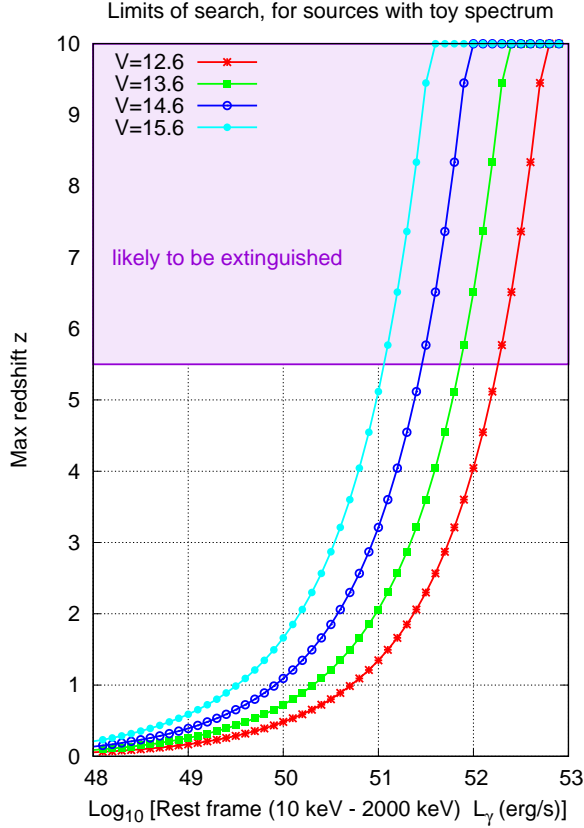}
   \end{center}
    \caption{Upper limits to the redshifts at which our survey
                could detect prompt optical flashes from GRBs.}
    \label{fig:limitz}
\end{figure}

When we set the value of $A$ so that the model
optimally matches the gamma-ray and optical observations
of the three comparison events
(to within an order of magnitude in all cases),
we derive a value of
$L_\gamma = 6 \times 10^{52}$ erg/s.
This implies that events similar to those in the comparison set
would be detected to redshifts $z > 9$ 
even for our brightest choice of magnitude limit.

Given a redshift, 
one can compute the comoving volume within a spherical region
and convert the areal rates listed in Table
{\ref{tab:rates}}
into volumetric form.
These rates, in units of events per cubic Gpc per year,
are displayed in Figure
{\ref{fig:limitvrate}}.
We arbitrarily set an upper limit of
$z = 10$ for the explosion of GRBs,
causing the rates to reach fixed values
at high luminosities, when the flashes
can be seen throughout the entire volume.
We list those fixed rates in
Table
{\ref{tab:volrates}}.
Note that when the search can detect events
anywhere in the volume, the rate for particular
limiting magnitude depends solely on 
the control time;
thus, the case of $V = 15.6$ ceases to provide
the most stringent upper limit because its control
time was smaller than that of $V = 14.6$ or $V = 13.6$.

\begin{figure}
   \begin{center}
      \includegraphics[width=80mm]{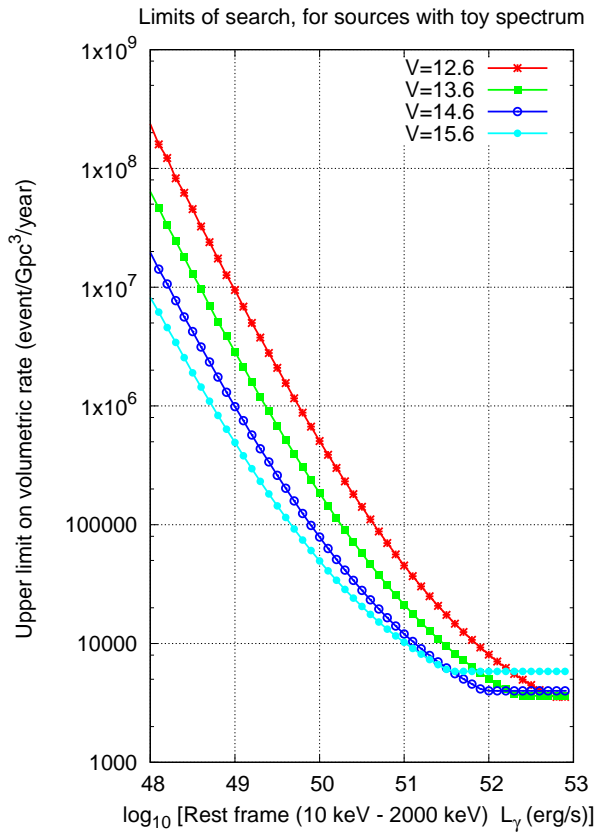}
   \end{center}
    \caption{Upper limits to the volumetric rates of GRBs with
                bright prompt optical flashes.}
    \label{fig:limitvrate}
\end{figure}

\begin{table}
  \caption{Upper limits (95\% confidence level) for rates of very luminous sources}
  \label{tab:volrates}
  \begin{tabular}{lllll}
    \hfil \phantom{if} & $V=12.6$ & $V=13.6$ & $V=14.6$ & $V=15.6$ \\
    \hline
     if $L_\gamma$ (erg/s) is greater than  & $6.3 \times 10^{52}$ & $2.5 \times 10^{52}$ & $1.0 \times 10^{52}$ & $4.0 \times 10^{51}$ \\
     then rate (${\rm Gpc^{-3} \thinspace yr^{-1}}$) is less than    & $3600 $ & $3700$  & $4000$ & $5800$ \\
  \end{tabular}
\end{table}

Let us briefly consider one implication of these calculations.
While the particular GRBs upon which our model was created
are due to events within the core of a massive star,
the separate class of short GRBs 
is thought to be due,
in part at least,
to the mergers of binary neutron stars (NS).
The rate of NS-NS mergers has become a hot topic
of late, thanks to the success of LIGO and Virgo.
\cite{2018arXiv181109296C}
provides a nice summary of both observational
measurements of the rate,
and theoretical expectations based on population 
synthesis and binary evolution.
The observations suggest somewhat higher rates
than the models,
with some estimates reaching a few thousand
per ${\rm Gpc^{3}}$ per year.
If (and we stress that this is a big qualification)
the NS-NS merger mechanism should produce a 
prompt spectrum similar to that of the long GRBs in our comparison set,
then continuing a search like ours for several months
or years could place meaningful constraints
on the rates of such mergers.

\section{Comparison with other transient searches}
\label{sec:comparison}

Many groups have examined the sky for ephemeral sources
over the past century,
but the great majority considered timescales 
much longer than the range of 1.5 - 11.5 seconds
that we explore in this work.
For example, the Deep Lens Survey Transient Search
\citep{2004ApJ...611..418B},
and efforts using data from 
ROTSE-III
\citep{2005ApJ...631.1032R},
MASTER
\citep{2007ARep...51.1004L},
DECam 
\citep{2019arXiv190311083A},
and PanSTARRS
\citep{2013ApJ...779...18B}
all required a source to remain bright for long
enough to appear in at least two images of moderate to long exposure time,
leading to minimum durations of hundreds or 
thousands of seconds.
Because all these projects could detect sources
much fainter than Tomo-e PM,
the control volumes
they examined were three to five orders
of magnitude greater than that described in this work.
However, since the transients we seek herein
would appear in only a single exposure
of all those projects, and so be discarded,
there seems little reason to consider them further.

Instead,
let us compare our survey briefly with 
three other projects which are sensitive to 
transients within our chosen timescale
and appear to be ongoing:
Pi of the Sky, Mini-Mega-TORTORA, and Gaia.

Pi of the Sky 
\citep{2008PhDT.......442S, 2014RMxAC..45..118O}
consists of one system in Atacama, Chile,
with two cameras,
and three systems in INTA, Spain, with four cameras each.
Each camera consists of a 85-mm f/1.2 camera lens
which focuses light onto a CCD, yielding a field
of $20^{\circ} \times 20^{\circ}$ with pixels subtending
$36^{''}$ on a side.
Exposures of length 10 seconds detect objects
down to $V \sim 12$;
this corresponds to a limit of about $V \sim 9.5$ 
in a 1-second exposure.
The weather at all sites is very good, providing
images of very large areas of the sky on many nights per year.

Mini-Mega-TORTORA, hereafter denoted MMT-9
\citep{2016RMxAC..48...91K},
is based at the Special Astrophysical Observatory in
the Caucasus, where the sky is somewhat cloudy.
Nine independent units, each built around an
85-mm f/1.2 camera lens 
and Andor Neo CMOS detector, 
provide images covering 
$9^{\circ} \times 11^{\circ}$ with pixels subtending
$16^{''}$ on a side.
Images with exposure times of 0.128 seconds are taken
7.5 times per second, providing very high time resolution.
The limiting magnitude in a single exposure is $V \sim 11$,
corresponding to a 1-second limit of $V \sim 13.3$.

The Gaia spacecraft rotates on its axis with a
period 108 minutes,
causing objects to drift across its focal plane.
\citep{2018MNRAS.473.3854W}
describe a technique for examining the measurements
from individual chips,
yielding sets of 11 exposures, each 4.5 seconds in length,
covering a span of 45 - 50 seconds.
Stars to $G \sim 20.5$ can be detected in each chip,
equivalent to $V \sim 19.0$ in a 1-second exposure.
Over the course of one day, the spacecraft
will scan roughly 1000 square degrees of the sky.
Due to the way in which objects are detected and followed
across the Gaia focal plane,
transients with the very short durations (1.5 - 11.5 seconds)
we consider in this paper
will be detected with a low efficiency ($\sim 0.1$)
and appear in at most two of the nine Astrometric Field chips.
Transients of this sort were not included in the 
analysis of
\cite{2018MNRAS.473.3854W},
nor do they appear in Gaia DR2,
but it is possible that researchers might make special efforts
to identify them in the low-level data products.

How do these projects compare to Tomo-e 
for finding transients with durations of order one second?
We will compute a statistic,
``relative control volume (RCV),''
which accounts for the properties of each survey,
yielding a relative measure of the volume
monitored per day;
larger values of RCV correspond to more frequent
observations of transient objects,
and thus more stringent limits on their rates.
We use a simple Euclidean model in these calculations.

$$
{\rm RCV} = \left( {{\rm hours\ active}\over{\rm day}} \right)
         \left( {{\rm sq. deg.}\over{\rm hour}} \right) 
         \left( d_{\rm max}^{\thinspace 3} \right) 
$$

The depth $d_{\rm max}$ of each survey is measured relative
to that of Tomo-e PM, based on the limiting magnitude
for an equivalent 1-second exposure.
We have estimated an average duration of 8 hours 
per night for the ground-based instruments,
and have assumed that both Pi of the Sky and MMT-9
are configured for maximum sky coverage.
We have multiplied the ``area scanned per hour''
entry for Gaia by $0.1$ to account for its low
efficiency of detecting very brief transients.

\begin{table}
  \caption{Comparison of some active survey instruments}
  \label{tab:rcv}
  \begin{tabular}{ld{3.2}d{5.1}d{6.2}d{7.2}d{9.1}}
    \hfil \phantom{if} & ${\rm Tomo-e PM}$ & ${\rm Tomo-e}$ & ${\rm Pi Sky}$ & ${\rm MMT-9}$ & ${\rm Gaia}$ \\
    \hline
     Hours per day  & 8 & 8 & 8 & 8 & 24 \\
     Area per hour (sq.deg) & 1.9 & 20.4 & 6400 & 6900 & 3.9 \\
     Pixel size (arcsec) & 1.2 & 1.2 & 36 & 16 & 0.18 \\
	  Lim. mag in image{\footnotemark[$*$]} & 17.8/14.9 & 17.8/14.9 & 12 & 11 & 20.6 \\
     Time resolution (s)  & 0.5 & 0.5 & 10 & 0.12 & 4.5 \\
     Lim. mag in 1-sec{\footnotemark[$*$]}  & 18.5/15.6 & 18.5/15.6 & 9.5 & 13.3 & 19.0 \\
     Normalized $d_{\rm max}$  & 1.0 & 1.0 & 0.060 & 0.42 & 4.8 \\
     RCV           &  15.2 & 163 & 11 & 2300 & 10,000 \\
    \hline
  \end{tabular}
  \begin{tabnote}
     \footnotemark[$*$] For Tomo-e, first value under optimal conditions; second value faintest limit used in control time calculations
  \end{tabnote}
\end{table}

The Tomo-e PM system appears comparable to the Pi of the Sky
by this metric, and far below the MMT-9 and Gaia instruments.
Observations by the full Tomo-e system, which have recently
started at the time of writing (June 2019), 
will increase its RCV by roughly a factor of ten, simply
by increasing the number of CMOS sensors from 8 to 84;
this increase places Tomo-e roughly halfway between
Pi of the Sky and MMT-9, in a logarithmic sense.

Of course, this metric is not the only way to evaluate the
performance of any of these surveys.  Each has its own
strengths and weaknesses, and each can and does provide
valuable information to those interested in transient
events.  
One of Tomo-e's strengths is its relatively high angular
resolution and uniform PSF across the field of view:
it suffers less from crowding and image distortions than 
Pi of the Sky and MMT-9, which may allow it to make
more accurate measurements of some events, or even to detect
some marginal events which might appear blended in
other systems.
On the other hand, its location in the mountains of central
Japan, while providing a dark sky, also delivers cloudy
weather more frequently than the sites of Pi of the Sky
and Gaia.

\section{Funding}
Our studies are
supported in part by the Japan Society for the Promotion of Science
(JSPS) Grants-in-Aid for Scientific
Research (KAKENHI) Grant Number JP25103502, JP26247074,
JP24103001, JP16H02158, JP16H06341, JP2905,
18H04575, JP18H01272, JP18H01261, and JP18K13599.
This research is also supported in part by the Japan Science and
Technology (JST) Agency's Precursory Research
for Embryonic Science and Technology (PRESTO),
the Research Center for the Early Universe (RESCEU),
of the School of Science at the University of Tokyo,
and the Optical and Near-infrared
Astronomy Inter-University Cooperation Program.

\begin{ack}
This research has made use of NASA's Astrophysics Data System.
We thank the Minor Planet Center for providing the Minor Planet Checker
and NEO Checker services to the community.
Ned Wright's cosmological calculator 
\citep{2006PASP..118.1711W}
was a great aid in our work.
\end{ack}

\end{document}